\documentstyle[epsfig]{mn}
\def\be{\begin{equation}}
\def\ee{\end{equation}}
\def\bea{\begin{eqnarray}}
\def\eea{\end{eqnarray}}
\def\etal{{et al.}\thinspace}

\title
[Universal Temperature Profile]
{Implications of the Universal Temperature Profile for Galaxy Clusters}

\author[Suparna Roychowdhury and Biman B. Nath]
{Suparna Roychowdhury and Biman B. Nath\\
Raman Research Institute, Bangalore 560080, India\\
(suparna@rri.res.in, biman@rri.res.in)
}
\date{17 February 2003} 

\pagerange{\pageref{firstpage}--\pageref{lastpage}} \pubyear{2003}

\begin{document}

\label{firstpage}

\maketitle

\begin{abstract}
{
We study the X-ray cluster gas density distribution in hydrostatic equilibrium 
using the universal temperature profile obtained from recent simulations 
involving only gravitational processes. If this temperature profile is an 
indicator of the influence of gravitational processes alone on the 
intracluster medium, then the comparison of various X-ray parameters expected 
from this profile and the observed data would point towards any additional 
physics that may be required. We compare the entropy at $0.1 R_{200}$ and 
$R_{500}$, the scaled entropy profile, the gas fraction at $0.3 R_{200}$ and 
the gas fraction profile with recent observations and discuss the implications 
of this temperature profile in light of these data. We find that the entropy 
imparted to the gas from gravitational processes alone is larger than 
previously thought. The entropy at $R_{500}$ for rich clusters is consistent 
with data, whereas the entropy at $0.1R_{200}$ is still less than the observed 
values. We also find that the gas fraction in the inner region of clusters, 
expected from gravitational processes alone, is smaller than previously 
thought but larger than the observed data. It does show a trend with the 
emission-weighted temperature ($\langle T \rangle$) as shown by data. We 
therefore find that the role of any additional non-gravitational process 
influencing the physical state of ICM would have to be revised in light of 
these findings.
}
\end{abstract}


\section{Introduction}

The formation of structures in the Universe is believed to be hierarchical, 
as primordial density fluctuations, amplified by gravity, collapse and merge 
to form progressively larger systems. This hierarchical development leads to 
the prediction of self-similar scalings between systems of different masses 
and at different epochs. 

X-ray observations of the intra-cluster medium (ICM) provide an ideal probe 
to test these self-similar scalings. Observations however present a picture 
that is at variance with predictions from these self-similar scalings. The 
X-ray luminosity of low temperature (poor) clusters fall below the 
self-similar expectations, and the surface brightness profiles of these 
clusters are seen to be shallower than those of richer clusters. It is also 
instructive to view this in terms of the entropy of the intra-cluster gas, 
which in self-similar scaling should increase in a very simple scaling with 
the mean temperature of virialized systems, whereas observations show that 
gas in low temperature clusters have larger entropy than expected. Another 
probe of non-gravitational processes is the gas fraction by mass in these 
clusters. Observations show that the inner regions of poor clusters have 
less gas (compared to the total mass within those radii) than those of rich 
clusters.

These differences between theoretical expectations and observations have 
led to the emergence of a number of theoretical ideas to increase the 
entropy of the ICM by some non-gravitational process, especially in low 
temperature clusters, involving heat input from supernovae (Valageas \& 
Silk 1999; Kravtsov \& Yepes 2000; Wu \etal 2000), quasar outflows (Bower 
1997; Loewenstein 2000; Voit \& Bryan 2001; Nath \& Roychowdhury 2001), 
gas cooling (Knight \& Ponman 1997; Bryan 2000; Voit \& Bryan 2001; Pearce 
\etal 2000; Muanwong \etal 2001; Wu \& Xue 2002a; Dav\'e \etal 2002) 
and accretion shocks (Tozzi \& Norman 2001; Dos Santos \& Dor\'e 2002; 
Babul \etal 2002), though it appears that there are problems with many of 
them (Ponman \etal 2003 and references therein; hereafter PSF03).

A recent result from a very high resolution numerical simulation (Loken 
\etal 2002) offers a fresh look at the scaling laws expected 
from gravitational interactions. Loken \etal (2002) found a universal 
temperature profile for the ICM, when scaled to the emission weighted
temperature, from the assumptions of pure gravitational evolution of the 
cluster gas, with no input from any additional non-gravitational processes of 
heating or cooling. Interestingly, this temperature profile is a good match 
to the observed universal temperature profile, leaving aside the very central 
regions ($r \le 0.1 R_{vir}$) (De Grandi \& Molendi 2002).

The interesting aspect of this temperature profile obtained from this 
simulation is that it has a core within which the profile is 
flatter than in the outer region. It has been earlier noted by Eke \etal 
(1998) that interactions between dark matter and baryons imparts some 
entropy to the gas providing it with a core (and making it deviate in the 
central regions from the self-similar expectations in which gas density is 
proportional to the dark matter density). The entropy imparted to the gas 
was, however, thought to be important only in the very central regions 
($r \ll 0.1 R_{vir}$),and observations of entropy at $\sim 0.1 R_{vir}$ was 
thought to reflect other, non-gravitational influences, if there were any.

It is then important to determine the gas density profile corresponding to 
this temperature profile, assuming hydrostatic equilibrium, and determine 
the level of entropy obtained {\it only} from gravitational interactions, 
{\it without} the aid of non-gravitational processes. In this paper, we 
study various implications of this temperature profile, especially in light 
of the entropy problem mentioned earlier.

As we describe below, we find that the entropy imparted to the gas by 
gravitational interactions alone is larger than previously thought for 
poor clusters, {\it even at radii exceeding $0.1 R_{200}$}, and this is the 
main result of the present paper. We begin with the description of the 
background dark matter density profile that is assumed (\S 2). We then 
determine the gas density profile assuming hydrostatic equilibrium in 
\S 3 and discuss various aspects of our result, including comparison with 
relevant data in \S 4. We end with a brief discussion of our results in \S 5.

We assume throughout the paper $\Omega_{\Lambda} \, = \, 0.71$, $\Omega_0 \,
=\,0.29$, $\Omega_b \,={\,0.047}$ and $h\,=\,0.71$ which are the best fit 
parameter from WMAP (Spergel \etal 2003).

\section{Universal dark matter density profile}

The dark matter density profile, $\rho_{dm}(r)$ suggested by many high 
resolution $N$ - body simulations is well described by a self-similar form. 
We assume that the gas mass is negligible compared to the total dark 
matter mass and adopt this universal density profile for dark matter in 
clusters. The profile is expressed in terms of a characteristic radius 
$r_s$ (e.g., in Komatsu \& Seljak 2002):

\be
\rho_{dm} = \rho_s y_{dm}(r/r_s)
\label{eq:dm}
\ee
where $\rho_s$ is a normalization factor which represents a characteristic 
density at a characteristic radius, $r \, = \, r_s$. This characteristic 
radius describes a typical scale at which the profile slope changes from the 
outer value to the inner value. The functional form of $y_{dm}\,(x)$ is given 
by
\be
y_{dm}(x) = {1 \over x^{\alpha}(1+x)^{3-{\alpha}}}
\label{eq:ydm}
\ee

Here the parameter $\alpha$ characterizes the shape of the profile. Since 
the dark matter density profile is self-similar, the dark matter mass 
profile is also self-similar. So, the dark matter mass enclosed within a 
radius $r$ is
\be
M(\le r) = 4 \pi \rho_s r_s^3 m(r/r_s)
\label{eq:mass}
\ee
where, $m(x)$ is a non-dimensional mass profile given by

\be
m(x) = \int_0^x du \,u^2 y_{dm}(u) = \ln(1+x) - {x \over (1+x)};
\label{eq:int}
\ee

Here, the last equality is valid for $\alpha \,=\,1$ which is the much used 
NFW profile (Navarro \etal 1996, 1997), the integral being evaluated by 
Suto \etal (1998). 

The definition of the virial radius, $R_{vir}$, is the radius within which 
the total dark matter mass is confined, i.e., $M_{vir}\equiv M(\le c)$, 
where
\be
c \equiv {R_{vir}\over R_{s}}
\label{eq:cp}
\ee
is a dimensionless parameter called the 'concentration parameter'. 
Evaluating equation (\ref{eq:mass}) at the virial radius, the 
normalization factor, $\rho_s$, is fixed at;

\be
\rho_s = c^3 {M_{vir} \over 4 \pi R_{vir}^3 m(c)}
\label{eq:rhos}
\ee 

The virial radius, $R_{vir}\,(M_{vir},z)$ is calculated with the spherical 
collapse model (Peebles 1980), 

\be
R_{vir}=\Bigl [ {M_{vir} \over (4 \pi/3)\Delta_c(z)\rho_c(z)}\Bigr ] ^ {1/3} 
= \Bigl [ 
{M_{vir}c^3 \over 4\pi \rho_s m(c)} \Bigr ] ^ {1/3}
\label{eq:rvir}
\ee
where the second equality comes from evaluating $R_{vir}$ from equation 
(\ref{eq:rhos}). Here $\Delta_c(z)$ is the spherical over-density of the 
virialized halo within $R_{vir}$ at $z$, in units of the critical density 
of the universe at $z$, $\rho_c(z)$. Following Komatsu \& Seljak (2002), 
we assume a value $\Delta_c(z=0) = 100$ for a cosmological model with 
$\Omega_m = 0.29$ and $\Omega_{\Lambda} = 0.71$.

We follow Bullock \etal (2001) in adopting the approximation for $c$ as a 
function of the virial mass of the cluster. They give the median values of 
`c' and also the $1\sigma$ deviations:

\be
c=K\Bigl ({M_{vir} \over 1.5 \times 10^{13}h^{-1}M_{\odot}}\Bigr )^{-0.13}
\label{eq:cpfit}
\ee
with $K = 9$ reproducing the best-fit and $K = 13.5$ and $K=5.8$ reproducing 
the $+1\sigma$ and the $-1\sigma$ values in the concentration parameter.
These values of the concentration parameter are also consistent with
the findings of  Seljak \& Huffenberger (2003).

The above set of equations specify the dark matter density profile of a 
particular mass cluster. Next, we turn our attention to the density profile 
of the gas in hydrostatic equilibrium with this dark matter distribution.

To compare our results with observations, which usually uses the radius 
$R_{200}$ where the over-density is 200, we compute this radius in each case 
and present our results in the terms of $R_{200}$.

\section{Hydrostatic equilibrium of gas and dark matter}

Our aim in this section is to determine the density profile of the 
intra-cluster gas using the universal temperature profile (discussed later) 
and assuming that the gas is in hydrostatic equilibrium with the background 
dark matter potential. 

The typically smooth morphology of the X-ray emission from the hot 
intra-cluster medium leads naturally to the hypothesis that the gas is near 
equilibrium, stratified along isopotential surfaces in a mildly evolving 
distribution of dark matter, gas and galaxies. This suggests that 
the assumption of hydrostatic equilibrium for such relaxed clusters is mostly 
justified. 

\subsection{Universal Temperature Profile of Gas}

The ``universal temperature profile'' used for our calculation (Loken \etal 
2002) is (normalized by the emission-weighted temperature):

\be
{T \over \langle T \rangle} = {T_0 \over (1+{r\over a_x})^{\delta}}
\label{eq:temp}
\ee
where $\langle T \rangle$ is the emission-weighted temperature of the cluster, 
$T_0 = 1.33, \, a_x = R_{vir}/1.5$, and $\delta = 1.6$ on the radial range 
(0.04-1.0) $R_{vir}$. To determine the emission-weighted temperature
from the cluster mass, we use a relation that arises from adiabatic evolution
of the gas in cluster. Afshordi \& Cen (2002) have shown that the observations 
of Finoguenov \etal (2001) of $M_{500} \hbox{--}\langle T \rangle$ relation 
in clusters can be understood from gravitational processes alone. We therefore 
use this empirical relation ($M_{500} \hbox{--}\langle T \rangle$) derived by 
Finoguenov \etal 2001:

\be
M_{500}=(2.64^{+0.39}_{-0.34})10^{13} \, {\rm M}_{\odot} \Bigl ( {k_b \langle 
T \rangle 
\over 1 \, 
{\rm keV}} \Bigr 
)^{1.78^{+0.10}_{-0.09}}
\label{eq:M-T}
\ee
where $k_b$ is the Boltzmann constant and $M_{500}$ has been calculated 
self-consistently by taking the total mass within the radius where the
over-density is $\delta \ge 500$.

The main motivation of using this universal temperature profile is to see 
if this profile which arises just out of gravitational interactions alone 
in clusters predicts anything different from the previously used default 
temperature profiles from gravitational interactions alone. This temperature 
profile is a result of a high resolution simulation, without any 
input from non-gravitational processes, carried out by Loken \etal (2002) 
which makes use of a Eulerian-based, adaptive mesh-refinement code that 
captures the shocks that are essential for correctly modelling cluster 
temperatures (for details of the simulation, refer to Loken \etal (2002)). 
The temperature profiles of the simulated $\Lambda$CDM and SCDM clusters are 
remarkably similar and are well fit by this universal temperature profile. 

Finoguenov \etal 2001 also point out that the $M_{500} \hbox{--} \langle 
T \rangle$ relation becomes flatter ($M_{500} \propto T^{1.58}$) when low 
mass clusters ($M_{500} \le 5\times 10^{13} M_{\odot}$) are excluded i.e. 
the $M_{500} \hbox{--} T$ relation becomes closer to self-similar relation 
of $M_{500} \propto T^{1.5}$. We have tried out with a flatter $M_{500} 
\hbox{--} T$ relation and seen that the results do not change appreciably. 
The code used by Loken \etal (2002) was tested as a part of the Santa Barbara 
cluster comparison project (Frenk \etal 1999) in which 12 groups
simulated a Coma-like cluster using a variety of codes and resolutions. 
They have shown that their results are among the highest-resolution results 
presented in the paper(central resolution of $7.8h^{-1} kpc$). They have 
also pointed out that their results for the Santa Barbara cluster are in 
excellent agreement with the results obtained from those of a new, 
completely independent code (Kravtsov, Klypin \& Hoffman 2002). 

This profile is in good agreement with the observational results of 
Markevitch \etal 1998 but diverges, primarily in the innermost regions, 
from their fit which assumes a polytropic equation of state. This profile is 
also in very good agreement with a recent sample of 21 clusters observed by 
{\it BeppoSAX} (De Grandi \& Molendi 2002) with and without cooling flows. 
Although the simulation result is consistent with the data at outer radii, 
there is some difference in the inner region of clusters, indicating that 
there could be some additional physics at small radii ($r <  0.1R_{vir}$) 
(Nath 2003). 

We emphasize that this temperature profile does not include the effects of 
cooling and galaxy feedback or for that matter any additional physics.
This temperature profile can be used, therefore, to probe the influence of 
gravitational interactions on the ICM, at radii $\ge 0.1 R_{vir}$.

It is instructive to compare the above mentioned temperature profile with 
the temperature profile calculated by assuming that $\rho_{gas}(r) = 
f_{gas} \rho_{dm}
(r)$, $\forall r$ where $f_{gas}=0.105$ (e.g in Bryan 
2000, Wu \& Xue 2002b) for a range of cluster masses. This is the 
self-similar model that has been used as a calibrator for the influence 
of gravitational
processes. Although one does not expect in reality for the above 
proportionality to hold for arbitrarily small radii, it has been expected 
that the influences of shocks resulting from gravitational interactions alone 
does not extend beyond $\sim 0.1 R_{vir}$ (e.g., Bryan 2000). This model 
therefore has been widely used to calculate the expected entropy of the gas 
at radii $\ge 0.1 R_{vir}$ from gravitational interactions alone, and compare 
this expectations with the observed values. 

We compare the temperature profiles assumed for the present model, and the 
profiles obtained using the self-similar model ($\rho_{gas} (r) \propto 
\rho_{dm}(r)$) in Figure 1, normalizing the temperature profiles by $T_{200}$,
where

\be
T_{200} = {GM_{200}\mu m_p \over 2R_{200}}
\label{eq:t200}
\ee
where,
$M_{200} = \int_0^{R_{200}}4\pi \rho_{dm} r^2 dr
$.

\begin{figure}
\centering
\includegraphics[width=85mm]{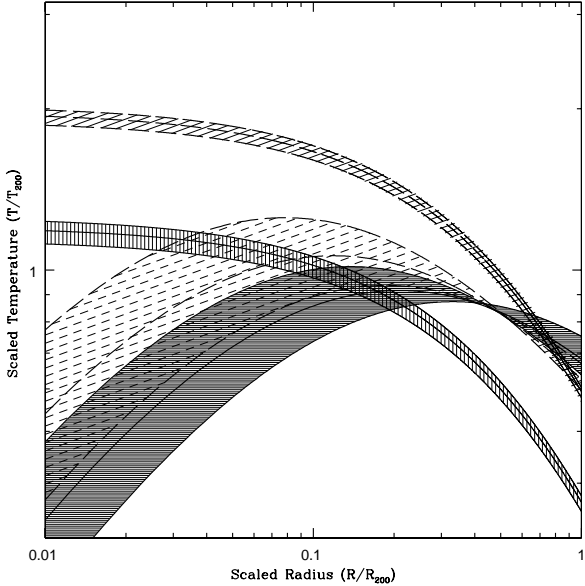}
\caption{The variation of gas temperature (scaled by $T_{200}$) with scaled 
radius for 2 clusters of different masses. The region shaded with vertical 
lines is the spread in gas temperature profiles of Loken \etal (2002) due to 
$1 \sigma$ spread in the concentration parameter `c' for a cluster of 
$\langle T \rangle$ 8.5 keV with the solid line in the middle being the 
profile for the best fit value of `c'. The region shaded with solid slanted 
lines is the universal temperature profile with the same spread in `c' for a 
poor cluster of $\langle T \rangle$ 0.85 keV with the long-dashed line in the 
middle being for the best fit value of `c'.  The region shaded dark with 
closely spaced horizontal lines is the result of the self-similar calculations 
for a cluster $\langle T \rangle$ 8.5 keV representing the spread in gas 
temperature for a $1 \sigma$ spread in `c' with the solid line in the middle 
being the result for the best fit `c'. Finally the region shaded with broken 
dashed lines is the gas temperature for the self-similar model of a cluster of 
$\langle T \rangle$ 0.85 keV with the spread being due to a spread in `c'.}
\end{figure}

\begin{figure}
\centering
\includegraphics[width=85mm]{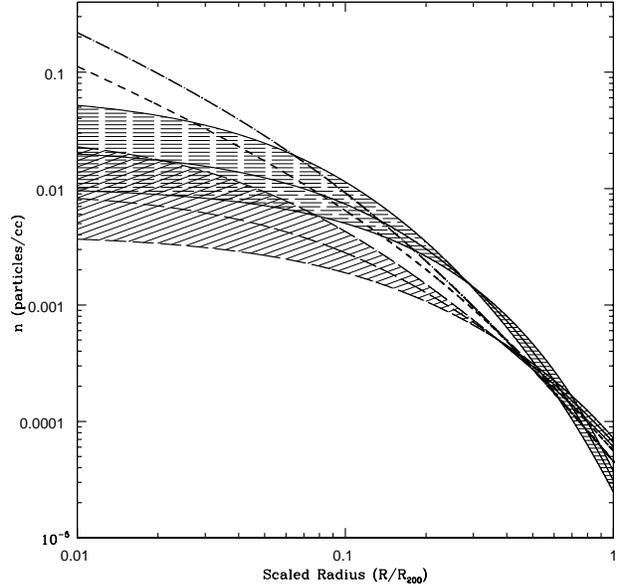}
\caption{Gas density profiles for 2 different clusters with scaled radius. 
The region shaded with horizontal lines is the spread in gas density as a 
result of the $1 \sigma$ spread in the concentration parameter for a rich 
cluster of $\langle T \rangle = 8.5$ keV. The region shaded with slanted 
lines corresponds to a low mass cluster (0.85 keV). The higher set of 
curves correspond to the self-similar model ($\rho_{gas}=f_{gas} \rho_{dm}$). 
The dot-dot-dashed line represents the cooler cluster (0.85 keV) and the 
dashed line the hotter of the two (8.5 keV).
}  
\end{figure}

It is seen that the universal temperature profiles (see Figure 1) 
flatten towards the inner regions of the cluster (within $0.2R_{200}$), 
whereas the temperature profiles calculated using $\rho_{gas}(r) \propto 
\rho_{dm}(r)$ (hereafter, referred to as self-similar profiles) dip at the 
central region of the cluster. 

\subsection{Density Profile of Gas}

We next numerically evaluate the density profile of the intra-cluster gas 
with the above defined temperature profile given in equation (\ref{eq:temp}) 
assuming hydrostatic equilibrium.

Let us consider a spherical gas cloud with temperature $T(r)$; then its 
density distribution $\rho_g(r)$ in hydrostatic equilibrium satisfies  
\be
{1 \over \rho_{gas}(r)}{d \over dr} (P_{gas}(r)) = - {GM(\le r) \over r^2}
\label{eq:hydro}
\ee
where,
\be
P_{gas} = \Bigl ({\rho_{gas}(r) \over {\mu m_p}}\Bigr )k_bT(r) 
\label{eq:pressure}
\ee
where $M( \le r)$ is the total mass inside radius $r$ (equation 
\ref{eq:mass}) 
and $\mu$ and $m_p$ denote the mean molecular weight (we use $\mu \, = \, 
0.59$) and the proton mass. The boundary condition which is imposed in 
evaluating this integral is that the gas-fraction, 
$f_{gas}$, within the virial radius is universal and is equal to 
$0.105$, as recently found by Ettori (2003) for a sample of low and high
redshift clusters. Since the total gas mass is negligible compared to the 
dark matter, $M_{total} \approx M_{dm}$, mass in dark matter, and therefore 
$f_{gas} \approx {M_{gas} \over M_{dm}}$.
The question whether or not $f_{gas}$ is independent 
of cluster mass has been a topic of debate in the literature (White \& 
Fabian, 1995; Ettori \& Fabian, 1999; Markevitch \etal 1999; 
Wu \& Xue, 2000). Although recent observations by Sanderson \etal (2003) 
show an apparent trend of $f_{gas}$ being smaller for lower temperature systems,
they also found that a universal value of $f_{gas}$ can fit their data. 
According to them, the unweighted mean of their data set of gas-fraction within 
the virial radius is a constant close to $f_{gas}=0.13 \pm 
0.01$ to $0.1 \pm 0.01$, the variation being due to the variation in the 
slope of the $M \hbox{--} \langle T \rangle$ relation. 

We note that this normalization provides a conservative estimate of the 
entropy of the gas from gravitational collapse alone since a lower value 
would only increase the entropy at all radii.
 
Figure 2 shows the gas-density profiles of 2 clusters of different masses. 
It compares the density profiles of the present model (lower set of curves) 
to the density profiles from self-similar assumptions (higher set of curves) 
with a constant proportionality factor of $f_{gas} = 0.105$ 
i.e., $\rho_{gas} = f_{gas}\rho_{dm}$. As expected, it is seen that the gas 
density is much shallower in the inner parts of the cluster as a result 
of the universal temperature profile being flat at the inner regions. 
Interestingly enough, the density profiles obtained from the universal 
temperature profile deviates from self-similar expectations at radii much 
larger than $0.1 R_{200}$. For a poor cluster with emission weighted 
temperature of $0.85$ keV the deviations become significant even at 
$r \sim 0.4 R_{200}$. 

We note that the emergence of a core in the gas density distribution has 
been noticed by previous authors of numerical simulations, even in the 
absence of non-gravitational heating and cooling processes (Frenk 
\etal 1999). This appears to result from the transfer of energy between 
baryonic and dark matter during merger events (Eke \etal 1998).

We have also done all the calculations described here for a coma-like 
cluster ($M_{vir} =  1.1 \times 10^{15} M_{\odot}$) for a S-CDM universe 
using the same parameters as in the Santa Barbara Cluster Comparison 
Project (Frenk \etal 1999) to check for the consistency of our method and 
results. We find that all the properties like gas-density profile, X-ray 
luminosity calculated using the prescription described in this paper 
match the simulated results of Frenk \etal (1999).

\section{Implications of the universal temperature profile and the derived 
gas density profile}

In this section, we focus on the implications of the above temperature and 
gas-density profiles on the other physical properties of this intra-cluster 
gas like entropy, gas-mass, variation of $M_{gas}$ with cluster mass or 
emission-weighted temperature, $T$ and gas-fraction, $f_{gas}$ in light of 
recent observations.
  
\subsection{Entropy Profiles and Scaling Properties}

For convenience, 'entropy' for the intra-cluster gas is defined as 

\be
S \equiv {T \over n_e^{2/3}} \, ({\rm keV \, cm^2})
\label{eq:ent}
\ee
where, $T$ is the temperature of the gas and $n_e$ the particle density. 
This quantity is directly related to observations. This has been referred 
to by a number of authors as 'adiabat', since (apart from a constant 
relating to mean particle mass) it is the coefficient relating pressure 
and density in the adiabatic relationship $P = K\rho^{\gamma}$. Hence $S$ 
is conserved in any adiabatic process. Note that the true thermodynamic 
entropy is related to this definition via a logarithm and additive constant.

In this section, we discuss the scaled entropy profiles (scaled with the 
emission-weighted temperature, $\langle T \rangle$) obtained from our 
density and temperature profiles for 5 different mass clusters. We calculate 
the emission-weighted temperature corresponding to the profiles discussed 
above within a fiducial radius of $0.3R_{200}$ (in the band $0.5 - 10$ keV), 
using the Raymond Smith code, for a metallicity of $Z/Z_{\odot} = 0.3$:
\be
\langle T \rangle = {\int_0^{0.3R_{200}} 4\pi r^2n_i(r)n_e(r)\epsilon_
{0.5-10}T(r)dr \over 
\int_0^{0.3R_{200}} 
4 \pi r^2n_i(r)n_e(r)\epsilon_{0.5-10}dr}
\label{eq:T_em}
\ee
where $n_i$ and $n_e$ represent the ion and electron density and 
$\epsilon_{0.5-10}$ denotes the emissivity relevant for the $0.5-10$ 
keV band. We find that the emission weighted temperature obtained in this 
manner matches well (within $0.5 \%$) with the value assumed to 
calculate the temperature profile itself (from equation \ref{eq:temp}). 
This shows that the system of equations used for our calculations is 
self-consistent.

Under the assumption that all these systems form at the same redshift, 
their mean mass densities should be identical. Hence in the simple 
self-similar case, where all have similar profiles and identical 
gas-fractions, $S$ will simply scale with emission-weighted temperature 
$\langle T \rangle$. We apply this scaling and scale the radial coordinate 
to $R_{200}$ for each system, derived as mentioned above.

\begin{figure}
\centering 
\includegraphics[width=85.0mm]{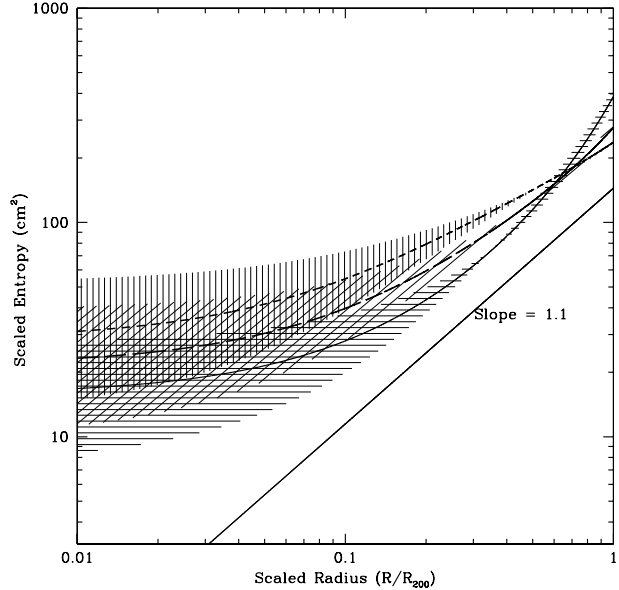} 
\caption{Scaled entropy profiles (scaled by $1/T_{200}$) for 2  
clusters. The region shaded with horizontal lines represents the spread 
in the scaled entropy profiles due to the $1 sigma$ spread in the 
concentration parameter `c' for a cluster with $\langle T \rangle$ of 
8.5 keV and the region shaded with slanted lines represents the spread in 
the scaled entropy profiles for a cluster with $\langle T \rangle$ of 
2.44 keV. The region shaded with vertical lines is for a low mass 
cluster of $\langle T \rangle$ 0.85 keV. The bottom line 
(solid) shows the slope of 1.1 expected from shock heating. Its 
normalization is arbitrary. 
}
\end{figure}

We show these scaled entropy profiles in Figure 3. It can be seen that the 
entropy profiles of the cooler systems, scaled in the above mentioned way 
tend to be significantly and systematically higher than that of rich 
clusters. In these derived entropy profiles, we notice that they generally 
flatten in the very interior parts of the clusters (inside $0.05R_{200}$) 
resulting from the flattening of density distribution at these radii. It is 
also seen that there is a noticeable tendency for the scaled entropy to be 
higher, at a given scaled radius, in cooler systems. Simulations and 
analytical models of cluster formation involving heating from accretion 
shocks, produce entropy profiles with logarithmic slopes of approximately 
1.1 (Tozzi \& Norman 2001), which agrees rather well with the slope of the 
calculated profiles outside $R \approx 0.2R_{200}$, for rich clusters but it 
does not show good agreement with the entropy profiles for poor clusters.

The general trend of our calculated scaled entropy profiles are  in good
agreement with the recent results of PSF03. However, it is seen that the 
values in general are systematically lower than that of PSF03 at around 
$0.01R_{200}$ and also at the outer reaches of the cluster.  

\begin{figure}
\centering
\includegraphics[width=65.0mm,angle=-90]{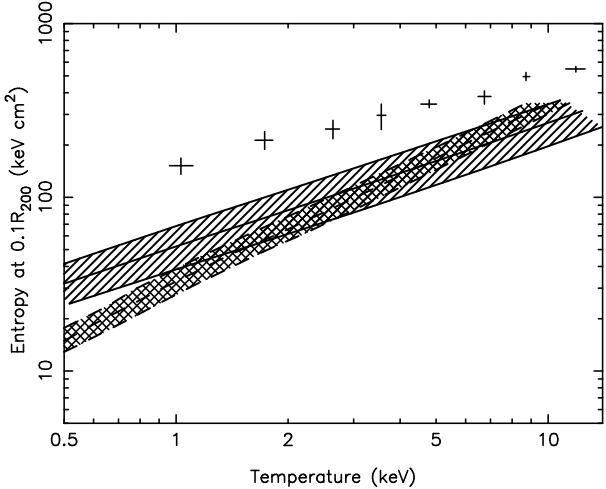}
\caption{Gas entropy at $0.1R_{200}$ as a function of emission-weighted 
temperature. The region shaded with slanted lines shows the spread in 
entropy due to the $1 \sigma$ spread in the concentration parameter with 
the solid line in the centre being the results of the best-fit `c' for the 
above described model, the data points are from PSF03 and the region shaded 
with crossed lines shows the spread in entropy due to the spread in `c' 
for the self-similar model ($\rho_{gas} \propto \rho_{dm}$) with the dashed 
line in the centre being the results of the best-fit `c'.}
\end{figure}

In Figure 4, we plot the variation of entropy $S$ at $0.1R_{200}$ with 
emission-weighted temperature, $\langle T \rangle$. The data points with 
the error bars are from PSF03. The region shaded with oblique lines 
represents the entropy calculated from the above described model with the 
$1 \sigma$ spread in the concentration parameter ``c' and  the 
region shaded with crossed lines corresponds to the self-similar density 
profiles. 
 
The discovery of an entropy floor in galaxy groups and clusters (Ponman 
\etal 1999) was based on the measurement of gas entropy at $0.1R_{200}$, 
in systems spanning a wide temperature range. This radius was chosen to 
lie close to the centre, where accretion shock-generated entropy should be 
minimum, hence maximising the sensitivity to any additional entropy, whilst 
lying outside the region where the cooling time is less than the age of the 
universe, and hence the entropy may be reduced. This initial study was 
improved by Lloyd-Davies \etal (2000) who derived an entropy floor value 
of $139 h^{-1/3}_{50} keV cm^2$ from a sample of 20 systems. However, the 
recent results of PSF03 show that there is no such entropy floor. 
They point out that an unweighted orthogonal fit to the data points, which 
have been grouped into temperature bins, gives a logarithmic slope of 
$0.57 \pm 0.04$, as opposed to $S \propto T$ in self-similar relation. 

As is seen in the figure, the entropy calculated from our model is higher  
than the previously calculated entropy from self-similar models  
(Wu \& Xue, 2002b) for the poor clusters and similar or slightly lower 
for rich clusters. However it is lower than the data points from PSF03, 
the difference being more pronounced for poor clusters. 
This is because of the fact that the density profile deviates from the 
self-similar models even at $0.1 R_{vir}$. 

\begin{figure}
\centering
\includegraphics[width=65.0mm,angle=-90]{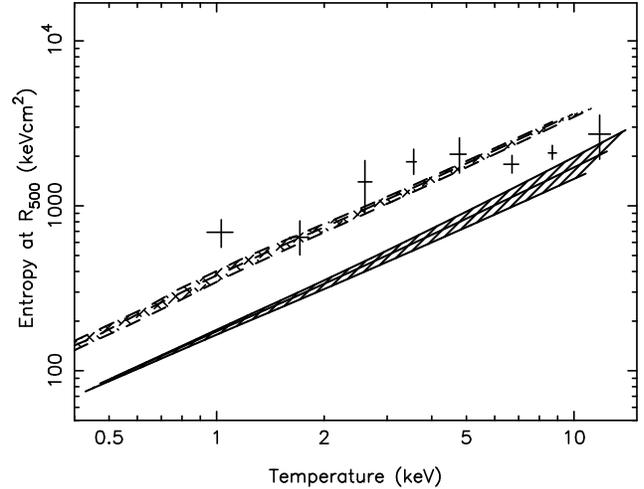}
\caption{Entropy at $R_{500}$ as a function of emission-weighted 
temperature. The solid line is the result of the above described model with 
the best-fit value of `c' and the region shaded with slanted lines being 
the spread in the entropy due to the $1 \sigma$ spread in `c'. The data 
points are from PSF03. The dashed line is the entropy calculated from 
the self-similar model($\rho_{gas} \propto \rho_{dm}$) with the best-fit 
value of `c' and the region shaded with crossed lines showing the spread 
in entropy due to the spread in `c'.
}
\end{figure}

We have noted earlier that the gas density profiles corresponding to the 
universal temperature profile is substantially flatter than the self-similar 
models even at radii larger than $0.1R_{200}$ (see Figure 2). 
It is therefore instructive to compare the entropy at larger radii with 
that from self-similar models and data. Recently Finoguenov \etal (2002) 
compared the entropy at $R_{500}$ (which is $\sim {2 \over 3} R_{200}$ for 
the range of cluster masses) with those expected from self-similar models 
and concluded that there is excess entropy even at this large radius,
indicating the large scale influence of preheating processes. 

In Figure 5, we show a plot of `entropy' $S (R_{500})$ with emission-
weighted temperature, $\langle T \rangle$. The region shaded with slanted 
lines represents the present model with the $1 \sigma$ spread in the 
concentration parameter and the region shaded with crossed lines shows the 
entropy calculated from the self-similar model ($\rho_{gas} (r)\propto 
\rho_{dm}(r)$). The solid line through the middle of the region shaded with 
slanted lines represents the results for the best-fit `c'. Data from 
PSF03 are also plotted for comparison.  
We find that the level of entropy at $R_{500}$ is reasonably consistent 
with
the observed values for rich clusters but they are lower than the observed 
values for intermediate and low mass clusters, with the deficiency becoming
appreciable
for poor clusters. We note that previous authors had re-normalized the 
expectations for self-similar models by matching them with the entropy of 
the richest clusters and thus concluded the presence of excess entropy at 
$R_{500}$. We do not normalize our calculated entropies in this manner in 
this paper. However, it still shows that there is a need for non-gravitational 
heating even at large radii especially for low mas clusters to fulfill 
the requirement of this excess entropy.       
\begin{figure}
\centering
\includegraphics[width=65.0mm,angle=-90]{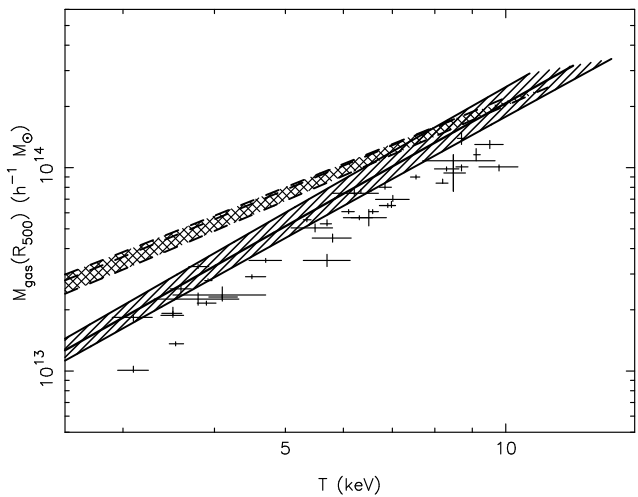}
\caption{Comparison of $M_{gas}\,(R_{500})-T$ relations. The data points 
with error bars represent gas mass determinations of MME99 within $R_{500}$. 
The solid line is the result of the present model using the best-fit value 
of `c' with the region shaded with slanted lines representing the spread 
in `c', the region shaded with crossed lines being the prediction of the 
self-similar model with the dashed line in the centre being the results of 
the best-fit value of `c'.  
}
\end{figure}

\begin{figure*}
\centering
\includegraphics[width=9.0cm,height=13.2cm,angle=-90]{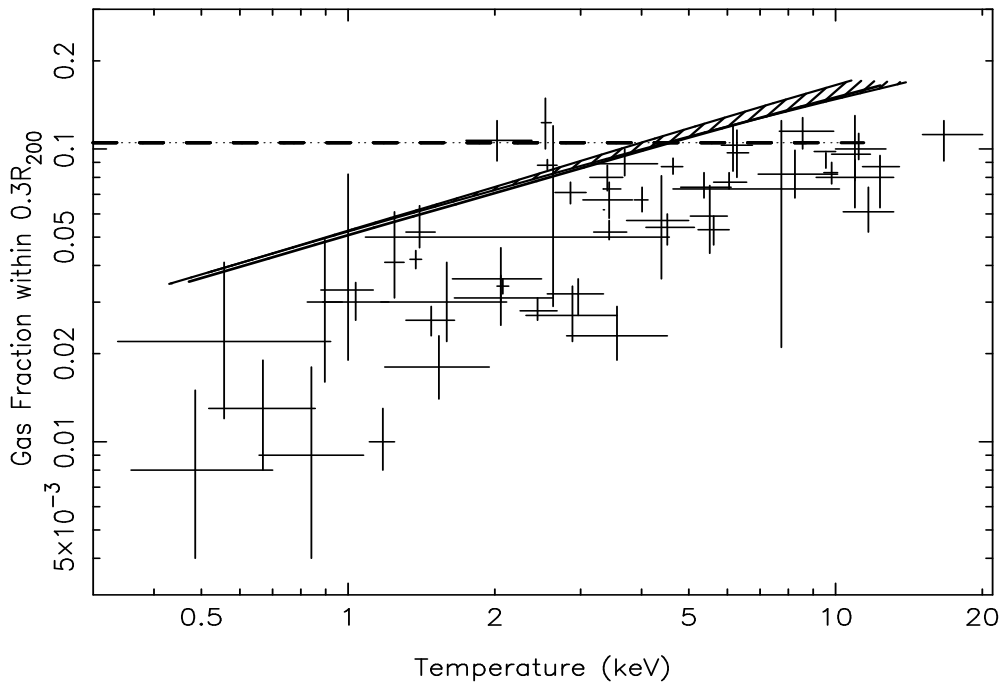}
\caption{Gas fraction within $0.3R_{200}$ as a function of emission-
weighted temperature,$\langle T \rangle$. The region shaded with slanted 
lines represent the above discussed model with spread in `c' with the 
solid line in the centre being the results of the best-fit value of `c' 
and the dashed line represents the model with $\rho_{gas} \propto 
\rho_{dm}$. The data points are from Sanderson \etal 2003.
}
\end{figure*}

The Figures 3 and 4 clearly show that the entropy of the ICM in the central 
regions ($\le 0.1R_{200}$) from gravitational processes alone, is larger 
than the previous expectations from self-similar models. It is then 
reasonable to conclude that the entropy imparted by interactions between 
dark matter and baryons (Eke \etal 1998) has been underestimated. 
Therefore, we find that gravitational interactions impart more entropy to 
the gas in the central regions than estimated earlier but the problem of 
excess entropy for low mass clusters still remains. It is interesting to 
find that the entropy at $R_{500}$ in the present model is  consistent 
with the observed values for rich clusters but are low for poorer clusters 
which probably confirms the requirement of some process which would help
preferentially increase their entropy. It is not surprising to find
consistency in the case of rich clusters though,
as we have already noted that the
temperature profile of Loken \etal (2002) is consistent with the
observed profiles at outer radii.

\subsection{Gas Distribution}
In this section, we discuss the effects of this temperature profile and the 
resulting density profile on the relation between $M_{gas}$ with mean 
emission-weighted temperature and the gas-fraction, $f_{gas}$ profiles and 
the variation of $f_{gas}$, in the inner regions, with emission-weighted 
temperatures.

\subsubsection{$M_{gas}$ ($R_{500}$) - $\langle T \rangle$} 

In figure 6, we present the $M_{gas}-\langle T \rangle$ relation as 
predicted by the present model (solid line), the relation derived from 
the self-similar model (dotted line), and the data points from Mohr \etal 
(1999) within $R_{500}$. It is seen that the gas mass within $R_{500}$ 
calculated from the present model is slightly higher than the data points, 
but lower than the expectations from self-similar models for clusters 
with $\langle T \rangle \le 3 $ keV and slightly higher for clusters 
with $\langle T \rangle \ge 3 $ keV, which was previously thought to be 
the result of gravitational processes alone. The logarithmic slope of our 
curve ($\sim 2.08$)is steeper than the self-similar slope of $1.5$, and 
close to the observed slope. 

\subsubsection{Gas fraction $f_{gas}$ and its spatial variation}  

The variation in the gas fraction ($f_{gas}$), evaluated within a 
characteristic radius of $0.3R_{200}$ is shown is Figure 7. The solid line 
is the calculated gas fraction from the present model and the dotted 
line shows the results of self-similar model. The data have been taken from 
PSF03. It is noted here that the gas fraction obtained from the present 
model is slightly higher than the data points. However, it is also seen 
here that there is a clear trend for cooler systems to have a smaller mass 
fraction of X-ray emitting gas in the central regions, as a result of 
gravitational processes alone. The gas fraction obtained from the self-
similar model is a constant as it should be by definition.

To understand the behaviour of gas fraction with radius better, we have 
plotted $f_{gas} (r)$ with the scaled radius, $r/R_{200}$ in Figure 8 for 
5 clusters of different masses. The general trend seen here is for 
gas-fraction to rise monotonically with radius (especially for clusters 
with $\langle T \rangle \le 3$ keV) all the way till $R_{200}$. There is,
however, a 
flattening of the profiles for the rich clusters ($\langle T \rangle \ge 3$ 
keV) beyond $0.5R_{200}$ with a slight bump around $0.3R_{200}$. It should
be noted that the fit provided by Loken \etal 2002 is accurate to about 
10 $\%$ for $r \le 0.5R_{vir}$ and underestimates their simulated 
temperature profile in this region. This may increase the gas density 
profile at smaller radii and account for the bump in Figure 8.
It can be seen clearly that the profiles lie in order of 
temperature such that, at a fixed radius, gas fraction decreases as 
temperature decreases, mirroring the trend in Figure 7.

If cluster evolution (with gravitational processes alone) were an entirely 
self-similar process, then $f_{gas}$ should be a constant at a given over-
density in all objects. Figure 9 clearly shows that this is not the case, 
even for evolution in the presence of gravity alone. Poor clusters seem to 
have a much lower gas mass fraction compared to rich clusters at the same 
over-density which was also seen by David \etal 1995. Again, a bump is seen 
for the richest cluster as seen in Figure 8. We have already discussed 
the probable reasons for this feature earlier.

\subsection{X-ray Luminosity-Temperature Relation}

\begin{figure}
\centering
\includegraphics[width=80.0mm]{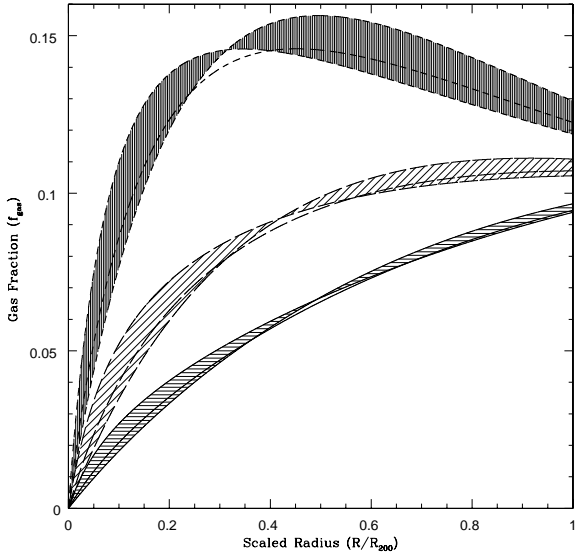}
\caption{Spatial variation of gas fraction within a given radius (normalized 
to $R_{200}$) for three different clusters with different emission-weighted 
temperature $T$. The solid line with the region shaded with horizontal 
lines represents the coolest system (0.85 keV), increasing in temperature 
through long-dashed with a region shaded with slanted lines (2.22 keV) and  
finally short-dashed (8.5 keV) line with the region shaded with vertical 
lines. These shaded regions represent the spread in gas fraction due to the 
$1 \sigma$ spread in `c'.
}
\end{figure}

In this section, we compute the bolometric X-ray luminosity, corresponding 
to the profiles discussed above (in the band 0.5-10 keV), using the 
Raymond-Smith code, for a metallicity of $Z/Z_{\odot}$ = 0.3. We compute 
the luminosities within the virial radius $0.3R_{200}$. The X-ray luminosity 
is computed as,
\be
L_X = \int_0^{0.3R_{200}} 4\pi r^2 n_i(r)n_e(r)\epsilon_{0.5-10}
\ee 

where $n_i$, $n_e$ represent the ion and electron density and 
$\epsilon_{0.5-10}$ denotes the emissivity relevant for 0.5-10 keV band. 
We present the results in Figure 10. 
\begin{figure}
\centering
\includegraphics[width=80.0mm]{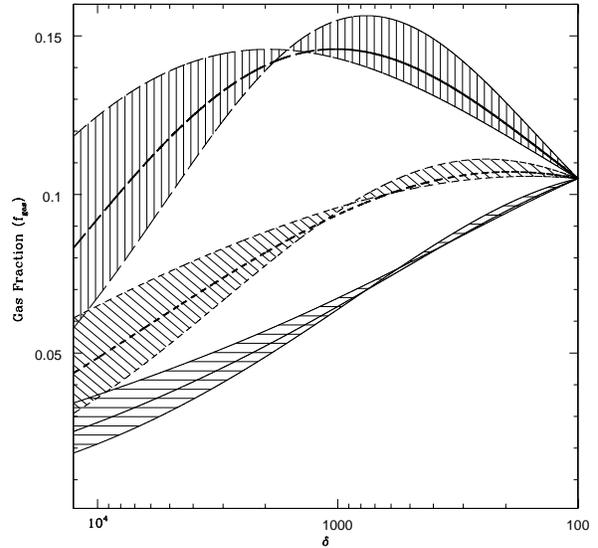}
\caption{Gas mass fraction as a function of the over-density $\delta$, 
for five different clusters. The solid line represents the coolest 
system (0.81 keV), increasing in temperature through dashed (1.26 keV), 
dot-dashed (2.22 keV), dotted (3.29 keV) and finally dash-dot-dot-dotted 
(7.36 keV) line.
}
\end{figure}

\begin{figure}
\centering
\includegraphics[width=65.0mm,angle=-90]{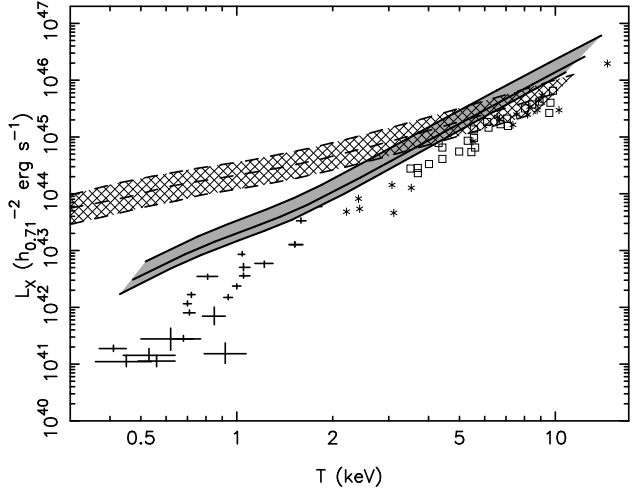}
\caption{Relation between bolometric X-ray luminosity $L_X$ and emission-
weighted temperature ($\langle T \rangle$). The data points represented 
by 'stars' show measurements of clusters with insignificant cooling flows 
compiled by Arnaud \& Evrard (1999). Open squares show cooling 
flow-corrected measurements by Markevitch \etal (1998). The data points 
with error bars show group data from Helsdon \& Ponman (2000). The shaded 
region represents X-ray luminosity calculated using the above model with the 
$1 \sigma$ spread in `c' with the solid line representing the median value 
and the region shaded with crossed lines is the result of the self-similar 
model with the same spread in `c'. The models assume a $\Lambda$CDM 
cosmology with $\Omega_M$ = 0.29, $\Omega_{\Lambda}$ = 0.71, and 
$\Omega_b$ = 0.047, and a Hubble parameter of $h$ = 0.71 has been applied 
to the models and the data. 
}
\end{figure}

It is seen, from Figure 10, that the luminosity calculated from the present 
model (with the above quoted values of the cosmological parameters) 
is close to the observed data (scaled with $h$ = 0.71 as used for the 
models) for clusters with emission-weighted temperature $\langle T 
\rangle$ above 1.0 keV. The data points are from Arnaud \& Evrard 1999 and 
Markevitch 1998 for clusters above 5.0 keV and from Heldson \& Ponman 2000 
for low $\langle T \rangle$ regime i.e. for low mass clusters and groups 
are also shown. Interestingly, compared to the dotted line which represents 
the self-similar model ($\rho_{gas} \propto \rho_{dm}$), the present model 
shows that the luminosity for low mass clusters (below 5 keV) is closer to 
data points. However, the luminosity is still somewhat
 over-estimated in this model 
when compared to data. This again indicates that there is requirement for 
some non-gravitational heating (preferentially in low mass clusters) to 
reduce the gas density further and thus the X-ray luminosity. We note here 
that the X-ray luminosity depends strongly on the assumed metallicity, 
especially for low temperature systems, and the uncertainty over abundance 
of gas in poor clusters is yet to be resolved (Buote 2000, Davis 1999).

In Figure 11, we have plotted the X-ray luminosity integrated within a 
radius $R_{1000}$ where the over-density $\delta \ge 1000$. The results 
of our calculations are compared with the best-fit results of Ettori 
\etal 2002. It is interesting to find that the X-ray 
luminosities calculated from the 
above model lie within the $1 \sigma$ spread of the results of
Ettori \etal 2002.

\begin{figure}
\centering
\includegraphics[width=65.0mm,angle=-90]{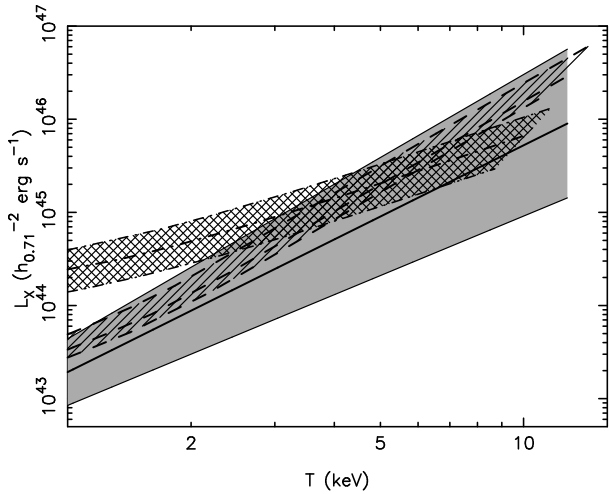}
\caption{Relation between bolometric X-ray luminosity $L_X$ and emission-
weighted temperature ($\langle T \rangle$). The shaded region enclosed in 
between the two solid lines represents the $1 \sigma$ spread which are the 
results of the best-fit analysis of the data of Ettori \etal 2002 with the 
solid line through the middle being the results of using the best fit 
parameters. The region shaded with slanted lines is the $1 \sigma$ spread 
calculated using the $1 \sigma$ deviations in the concentration parameter 
stated above with the dashed line being the results of the median value of 
`c' and the region shaded with crossed lines in the result of similar 
calculations using the self-similar model.
}
\end{figure}

\section{Discussion}

To recapitulate, our attempt here has been to study in detail the 
implications of the universal temperature profile obtained from recent 
high resolution simulations (Loken \etal 2002), with no input from 
non-gravitational processes, such as heating or cooling. We have 
compared the predictions of entropy and gas fraction from this temperature 
profile with the self-similar model ($\rho_{gas}\propto \rho_{dm}$), which 
has been used in the literature as a reference model for calibrating the 
influence of gravitational processes, at radii $r \ge 0.1 R_{vir}$. The 
only assumption that we have made in this paper is that of hydrostatic 
equilibrium, apart from the assumption of the background dark matter 
density profile. We have found that the result of gravitational processes 
alone is much different from what has been used in the literature so far. 

Firstly, the widely used assumption that entropy imparted to the gas from 
the interaction between dark matter and baryons is limited to the very 
central region, $r \ll 0.1 R_{vir}$ appears to be simplistic. The 
corresponding assumption that the gas density profile expected from 
gravitational processes alone is proportional to the dark matter density 
profile at $r \ge 0.1 R_{vir}$ seems to be  violated. Curves in Figure 2 
show that the gas density profile deviates from the simple proportionality 
at a much larger radii, $r \ge 0.5 R_{200}$, and this deviation is larger 
for lower temperature systems.

If the temperature profile obtained by Loken \etal (2002) is confirmed to 
be the one that is expected from gravitational processes alone, then the 
implications discussed here are inevitable. Instead of using the 
self-similar models, one must use the present model to benchmark the 
expectations from gravitational processes, and then compare it with the 
data to infer the need for any additional physics.

However, to conclude definitively about the presence of any 
non-gravitational heat in the ICM, one needs to know better about the dark 
matter profile in clusters and to resolve the uncertainties in the 
concentration parameter in dark matter halos of clusters. This idea was
also explored by Lloyd-Davies \etal (2002) who
concluded that the scatter in the concentration parameter plays a 
role in the determination of the contribution of non-gravitational 
heating in the intra-cluster medium. Also, in this context, it is 
worthwhile to point out that the dependence of `c' on redshift would have 
an effect on the results. This is because the assumption that the clusters 
were all formed at a redshift $z_f = 0$ is a simplification of the actual 
picture. Different clusters would form at different redshifts according to 
the theory of structure formation (Press \& Schechter, 1974) and that 
would mean that the concentration parameters would be different. The 
values of `c' would be lower as $c \propto {1 \over (1+z)}$ (as found by 
Bullock \etal 2001). This will make the results slightly different, e.g., 
entropy at $0.1R_{200}$ would be higher than the values plotted now and 
thus closer to the data and the X-ray luminosity would also be lower and 
thus closer to the data points. However, the entropy at $R_{500}$ will
increase as a result of this and the gas fraction at $0.3 R_{200}$ will
decrease, thus being closer to the respective data points.

We find that for entropy at $0.1 R_{200}$ the present model produces
lower entropy than the data for all clusters. The difference is larger for 
poorer clusters. The problem would then be to increase the entropy, 
especially for poor clusters, as has been required from theoretical models 
earlier. The requirement of non-gravitational heating as estimated from 
earlier theoretical models would be reduced because as it is seen here that 
the entropy at inner regions were under-estimated in earlier models. 
We also find that the entropy expected at $R_{500}$ is consistent  with
the data for rich clusters but it is lower than the data points for low
mass clusters emphasizing the need for some process to heat the gas even
at large radii.
It is however interesting that the scaled 
entropy profiles are similar to that observed 
(PSF03, Mushotzky \etal 2003). 

It is possible that the discrepancy between the present model and the data 
for entropy at $0.1R_{200}$ and X-ray luminosity for rich clusters can be 
alleviated by gaseous processes such as thermal conduction. As 
Loken \etal (2002) commented, the observed temperature profile is more 
flattened than their simulated profile, with somewhat larger temperature 
at $\sim$ $0.1R_{200}$. This could be due to thermal conduction 
(e.g., Nath 2003). This would then decrease the density 
in the inner regions and 
would (1) increase the entropy \& (2) decrease X-ray luminosity to be 
consistent with data for rich clusters. We note that thermal 
conduction is however less important poor clusters. 

We have noted that normalizing the gas content of clusters with a constant 
gas fraction for all clusters provides a conservative estimate of entropy 
from gravitational collapse alone. Also, we have found that changing the 
exponent of the $M_{500}\hbox{--} T$ relation to 1.5 does not change the 
results much.  

The total gas mass at a fiducial radius of $R_{500}$ expected from the 
present model are again closer to the data than the previous 
self-similar model. The gas fraction at $0.3R_{200}$ calculated from this 
model is higher than the data but it agrees better in comparison to the 
previously calculated ones from the self-similar model. It also
rises with the temperature of the cluster as observed.

There is, therefore, some difference between the expectations from the 
present model and the data. If the present model is a realistic indicator 
of gravitational processes alone, then the results from this model should 
be compared with the data, to determine the requirements of additional 
physics, if any, to explain the data. 

\section{Conclusion}

The primary aim of this work was to study the implications of the 
'universal temperature profile' arising out of pure gravitational 
interactions in galaxy clusters. We have determined the gas density 
profile corresponding to this temperature profile, and studied various 
implications of this profile. The only assumptions made in this paper is 
that of hydrostatic equilibrium and the temperature profile of Loken 
\etal 2002. 

We have also shown the dependence of the above results on the uncertainty 
in the knowledge of the concentration parameter `c'.

Given the uncertainty in the concentration parameter, the main results 
are summarized below:

(a) Gas density profiles expected from gravitational processes alone is 
flatter than previously thought, even at radii much larger than 
$0.1 R_{vir}$.

(b) Entropy expected from gravitational processes alone at $0.1 R_{200}$ 
are larger than previously thought, especially for low mass clusters, 
but still lower than the observed values, with the discrepancy increasing 
for low mass clusters. The entropy expected at $R_{500}$ is consistent
with observed values for rich clusters but it is lower than the 
data points for low mass clusters.
Thus it emphasizes the need for non-
gravitational heating preferentially for low mass clusters even at large 
radii.

(c) Gas fraction in the inner region ($0.3 R_{200}$) expected from 
gravitational process alone is much smaller than previously thought, 
and but slightly higher than the observed values.

We therefore conclude that if the temperature profile of Loken \etal 
(2002) is indeed the result of evolution of the intracluster gas 
involving gravitational processes alone, then the contribution of 
non-gravitational processes to the physics of ICM would have to be 
revised.

\bigskip\bigskip
\noindent
{\bf Acknowledgment} : 
We are grateful to Trevor Ponman for sharing with us his paper before 
publication which inspired this work and for his comments as the referee. 
We also thank Dipankar Bhattacharya and Stefano Ettori for useful discussion.

\label{lastpage}

\end{document}